\documentclass{iau_FM}
\usepackage{graphicx}

\title[AURIGA thick disks] %% give here short title %%
{The interplay of 
internal and external processes 
in the buildup of disk galaxies: thick-disk star formation histories 
in AURIGA simulations
}

\author[F. Pinna, D. Walo-Mart\'in \& R. J. J. Grand]   %% give here short author list %%
{Francesca Pinna$^1$, Daniel Walo-Mart\'in$^{2,3}$
 \and Robert J.J. Grand$^{2,3}$}

\affiliation{$^{1}$Max Planck Institute for Astronomy, Koenigstuhl 17, D-69117 Heidelberg, Germany\\email: {\tt pinna@mpia.de}\\
$^{2}$Instituto de Astrof\'isica de Canarias, Calle V\'ia L\'actea s/n, E-38205 La Laguna, Tenerife, Spain\\
$^{3}$Departamento de Astrof\'isica, Universidad de La Laguna, Av. del Astrof\'isico Francisco S\'anchez s/n, E-38206, La Laguna, Tenerife, Spain
}

\pubyear{2022}
\jname{IAU Symposium S373, Volume X} 
\editors{Piero Benvenuti, ed.}
\begin{document}

\maketitle

\begin{abstract}
Recent integral-field spectroscopy observations have revealed that thick- and thin-disk star-formation histories are regulated by the interplay of internal and external processes. 
We analyze stellar-population properties of 24 spiral galaxies from the AURIGA zoom-in cosmological simulations, to offer a more in-depth interpretation of observable properties. 
We present edge-on maps of stellar age, metallicity and [Mg/Fe] abundance, 
and we extract the star-formation and chemical-evolution histories of thin and thick disks. Both show signs of the interplay between internal chemical enrichment and gas and star accretion. Thick disks show particularly complex stellar populations, including an in-situ component, formed from both slowly enriched and accreted more pristine gas, and an additional significant fraction of ex-situ stars. 
\end{abstract}

\firstsection % if your document starts with a section,
\section{Introduction}
Spatially resolved stellar-population analyses of massive galactic disks allow us to draw the spatial and temporal distribution of star formation during the mass assembly of disk-dominated galaxies. 
Properties of thick disks, old, metal poor and enhanced in $\alpha$ elements, trace mostly the early stages of galaxy formation, while younger metal-rich thin disks tell us about later evolutionary phases (e.g., \cite[Yoachim \etal\ 2008]{Yoachim2008}, \cite[Gallart \etal\ 2019]{Gallart2019}). 
While we know that thin disks form mostly in situ, with its intense and extended star formation being fuelled by gas accretion, there is no common agreement on how thick disks form. 
Three main formation scenarios were proposed to explain their properties: 
their stars may have formed in situ, at high redshift, from turbulent gas 
\cite[(Brook \etal\ 2004)]{Brook2004}, 
in satellites that were later directly accreted \cite[(Abadi \etal\ 2003)]{Abadi2003}, or 
in a preexisting thinner disk which was later dynamically heated 
\cite[(e.g., Di Matteo \etal\ 2011)]{DiMatteo2011}. 

Observational studies of thick disks are generally based on edge-on galaxies, where fainter thick disks can be morphologically decomposed 
from the bright thin disks \cite[(e.g., Comer\'on \etal\ 2018)]{Comeron2018}. 
Most of the spectroscopic studies, which revealed some variety in thick-disk properties, 
have supported one or another of the mentioned formation scenarios \cite[(e.g., Yoachim \etal\ 2008]{Yoachim2008a}, \cite[Kasparova \etal\ 2016]{Kasparova2016}, 
\cite[Comer\'on \etal\ 2019)]{Comeron2019}.
On the other hand, resolved star-formation and chemical-evolution histories of thick disks, extracted from deep integral-field MUSE observations, have recently unveiled a combination of in-situ and ex-situ stellar populations, resulting from complex scenarios with different mechanisms at play 
\cite[(Pinna \etal\ 2019a]{Pinna2019a}, \cite[Pinna \etal\ 2019b]{Pinna2019b}, \cite[Martig \etal\ 2021)]{Martig2021}. 

Here we probe these complex scenarios with the use of numerical simulations. 

\section{Overview}
{\underline{\it Method}}. 
We used 24 Milky Way-mass galaxies from the AURIGA zoom-in cosmological simulation sample \cite[(Grand \etal\ 2017)]{Grand2017}. We projected the galaxies in an edge-on view, to allow a similar analysis to what is usually done in integral-field spectroscopy observations. We performed a Voronoi binning to ensure a similar number of star particles in each spatial bin. We used for the analysis a region of radius the optical radius $R_{opt}$ of the galaxy \cite[(Grand \etal\ 2017)]{Grand2017} and of height $h_{scale}$, the standard deviation of vertical positions of stars at $R_{opt}$. 
For each galaxy, we fitted vertical luminosity-density profiles, extracted in four radial bins, with two components associated with the thin and the thick disks, each one represented by a hyperbolic secant square function.
The fits gave us the average height, for each galaxy, at which the thick disk starts to dominate. 

{\underline{\it Stellar populations}}.
We mapped the stellar populations of the 24 galaxies and found a variety of ages, total metallicities ([M/H]) and [Mg/Fe] abundances in the thick disks, with average values respectively between 4 and 11~Gyr, -0.5~dex and solar values (0~dex), and between 0.15 and 0.23~dex (see Pinna et al., in prep.). Thick disks are in all cases older, more metal-poor and more [Mg/Fe] enhanced than thin disks. 
We extracted star-formation histories from the thin- and thick-disk-dominated regions, in terms of the mass fraction of stars of a specific age (following \cite[Pinna \etal\ 2019a]{Pinna2019a}, \cite[Pinna \etal\ 2019b]{Pinna2019b} and \cite[Martig \etal\ 2021)]{Martig2021}. We color coded age bins according to their average metallicity and [Mg/Fe] abundance, to interpret these plots in terms of chemical evolution. 
\begin{figure}[b]
 \vspace*{-0.4 cm}
 \includegraphics[width=5.4in,trim={2cm 0cm 0cm 0cm}]{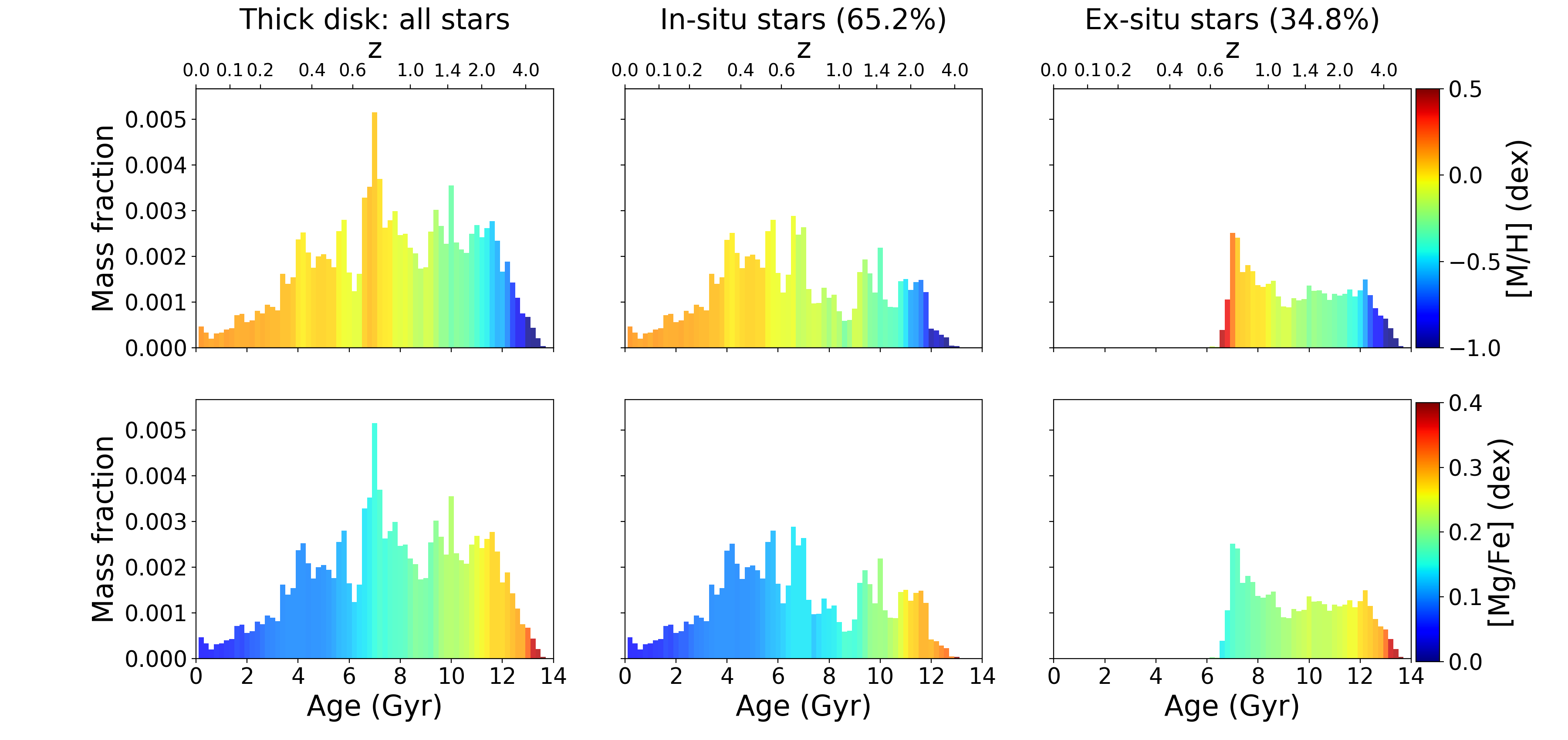} 
 \vspace*{-0.6 cm}
 \caption{Star-formation history of the thick disk in AU5, color-coded by total metallicity (top panels) and [Mg/Fe] abundance (bottom panels). Left panels include all stars, while middle and right panels include respectively only stars formed in situ or ex situ, indicating their corresponding thick-disk mass fraction on top of each column. }
   \label{fig3}
\end{figure}

{\underline{\it A representative galaxy: AU5}}. We show in Fig.~\ref{fig3} (left panels) the star-formation history of the thick disk in AU5, a barred spiral galaxy of stellar mass $\sim 7 \times 10^{10} M_{\odot}$ (see Pinna et al. in prep. for the full sample). At older ages, as expected from observations, we see a global increase of the star-formation rate (mass fraction in each age bin, from right to left), and a later decrease at young ages, with numerous star formation bursts across time. The colors display a global chemical enrichment, transitioning from metal-poor values and a strong [Mg/Fe] enhancement, to slightly subsolar metallicities and slightly supersolar [Mg/Fe] values. However, on top of the general trend, both [M/H] and [Mg/Fe] values oscillate at certain times across thick-disk evolution. 

{\underline{\it In-situ and ex-situ components}}. In order to understand the origin of these oscillations, which often correspond to star-formation bursts, we classified star particles as in situ or ex situ, by tracking them during the simulations. We mapped the mass fraction of accreted stars for the full galaxy sample, and found that they are mostly located in the thick-disk dominated regions (see maps in Pinna et al., in prep.). These regions host an accreted fraction between 7 and 61\% of their mass, with a median fraction of $\sim 28$\%. 

We decomposed the star-formation histories into in-situ and ex-situ components, and see that the oscillations in chemical abundances are driven by the combination of them. In AU5 (Fig.~\ref{fig3}, right panels), a satellite contributed $\sim 35$\% of the thick-disk mass. This satellite had its own chemical evolution, and stars with high metallicity were formed during the star-formation burst at the final stage of the merger (at ages of about 7~Gyr, probably involving also some gas stripped from AU5; note that these higher metallicities are still much lower than values in the thin disk of AU5). The in-situ component (middle panels) is characterized by several star-formation bursts with lower metallicity values (and slightly higher [Mg/Fe]) than slightly younger and older ages. These drops in metallicity are driven by the accretion of gas from outside the galaxy.

{\underline{\it Thick and thin disks in the full sample}}.
Although each galaxy in the sample has peculiarities, the results derived from AU5 are representative of most thick disks. For thin disks, as expected from previous observations \cite[(e.g., Gallart \etal\ 2019]{Gallart2019} and \cite[Martig \etal\ 2021)]{Martig2021}, we find a very low mass fraction of accreted stars, and the bulk of the mass was formed in recent times thanks to a large amount of accreted gas (for more details, see Pinna et al., in prep.).

\section{Implications}

{\underline{\it Thick-disk formation}}. 
These results affirm that although thick disks are mainly formed in situ (but with a strong contribution of gas accreted at the time of frequent mergers), they host a significant contribution of accreted stars. This accreted fraction varies from 7 to 61\%, on average 28\% of the mass.

{\underline{\it Galaxy evolution}}.
Star-formation histories of thick and thin disks, thus of disk galaxies, result from the interplay of internal and external processes, in which the internal chemical enrichment is challenged by the accretion of more pristine gas and more metal-poor stars.

\end{document}